\documentclass{svjour2}                    
\smartqed  
\usepackage{graphicx}
\begin{document}
\title{Quantum Non-Locality and Universe}
\author{P.Achuthan \and Narayanankutty Karuppath}
\institute{PAchuthan \at
Department of Mathematics,Indian Institute of Technology Madras,
            Chennai, Tamilnadu, India- 600 036 \\
           \email{achuthan@ettimadai.amrita.edu}   \\
           \\ \emph {Department of Mathematics, 
Amrita Vishwa Vidyapeetham (Deemed University), 
Coimbatore,Tamilnadu, India-641 105.} 
         \and Narayanankutty Karuppath\at Department of Physics, Amrita Vishwa Vidyapeetham,Coimbatore,Tamilnadu, India-641 105.\\}   
\maketitle
\begin{abstract}
The notion of Non-Locality (NL) in physics has great epistemological implications and impact on the perceptive of Reality. Numerous counter intuitive experimental results that clearly violate Bell's inequality have prompted several physicists to believe in 'mysterious' non-local correlations particularly in quantum systems of EPR type. Not commenting much on whether NL is true or not, our contention is that NL, in general, leads to the concept of holistic universe and further to a global concept of 'Consciousness-Centered-Cosmos' (CCC) as almost inevitable rational outcome. These suggestions though ostensibly unorthodox, are rather within rational and legitimate frame work of science. Our reasoning is quite distinct from the erstwhile doctrine of the role of consciousness in the measurement problem and state vector reduction although the latter may add further credence to the our contention. We address particularly the implications NL and discuss certain related aspects of new quantum world view in  general. The less recognized association of the principles of uncertainty with locality, causality, the arrow of time and reality are brought out. Perhaps these underpinnings are much less appreciated than necessary.

\keywords {Locality \and Reality \and Bell's inequality\and EPR\and Consciousness}
\end{abstract}
\section{Introduction}   
\label{intro}
The issue of Non-Locality (NL) is quite deep and enigmatic. It has much wider and deeper philosophical connotations than what one would normally envisage. NL has bearing on several aspects of physics including time and reality. Einstein referred to Quantum Non-Locality (QNL) as the 'spooky action at a distance'. QNL, as one justifiably feels, makes Quantum Mechanics (QM) much more inexplicable. The paramount importance of the issue of NL, particularly to physics can not be overstated. QNL, quantum reality, free will, determinism and such corresponding and related topics connected to the interpretation of QM have been subjects of debate for more than 80 years. These subjects are still being pursued in diverse contexts by many \cite{Hooft,HooftGod,Unni,Cramer}. Whether NL itself is really exits in physics or not may be deferred to a different occasion. For our immediate deliberations, we may take the actuality of NL or at least QNL to be true. Let us examine the epistemological implications of the concept of NL, as tacit by QM or otherwise.  Without renouncing the virtue of austerity and not be altogether speculative if one spans for more potential, one can make far reaching inferences. In this paper, we reason that NL in general and QNL in particular takes us to the idea of holistic nature of the universe that promote the model of Consciousness Centered Cosmos (CCC) coherently. CCC is one in which consciousness plays the role of a central agency defining the reality of the universe.
\section{Views Of Some Prominent Scientists}   
\paragraph{} Before directly taking up the actual problem, it would be instructive to have just a cursory look at what some of the present and old  noted physicists have said about NL, reality and consciousness. Below are only random and sparse sample of quotes taken from a very vast spectrum. 
Newton considered action at a distance to be an absurdity as can be seen from his writings, \textit{"That one body may act upon another at a distance through a vacuum, without the mediation of any thing else, by and through which their action and force may be conveyed from one to the other, is to me so great an absurdity, that I believe no man who has in philosophical matters a competent faculty of thinking, can ever fall into it."}-Sir Isaac Newton

The English translation of the original German version of Pauli's classical book on quantum mechanics says, \textit{"This solution (to the long-sought wave-particle duality problem) is obtained at the cost of abandoning the possibility of treating physical phenomena objectively .i.e. by abandoning the classical space-time and causal description of nature which essentially rests upon our ability to separate uniquely the observer and the observed}-Wolfgang Pauli (Translated from German, 1958 \cite{Pauli}). 
\textit{"Among the scientists and philosophers who have suggested a link between consciousness and Quantum theory are Alfred North Whitehead, Erwin Schr\"{o}dinger, John Von Neumann, Eugene Wigner, David Albert and Barry Loewer, Euan Squires, Evans Harris Walker, C.Stuart, Y.Takahashi, H.Umezawa, Amit Goswami, Aushalom Elitzur, Alexander Berezin, Roger Penrose, Michael Lockwood and John Eccles"}-Henry Stapp\cite{Stapp}.
\textit{"Throughout this book the central underlying theme has been the unbroken wholeness of the totality of existence as an undivided movement without borders"}-David Bohm \cite {Bohm}.\\ 
\textit{"The only justification for our concepts and system of concepts is that they serve to represent the complex of experiences; beyond this they have no legitimacy"}\cite {HooftDeterm}.
\textit{"The 'real external world' is the sum total of all experienced events about which subjects agree."} 

Some of these eminent thinkers have even worked on details of it also. Our focus is mainly on one of the exceptionally important aspects, namely NL. Our arguments about CCC is distinct from that of earlier proposal of the mechanism of state vector reduction (or wave length collapse) by the measurement chain leading to consciousness by von Neumann and the later people.
\section{Mach's Principle}    
Perhaps Mach's Principle can be seen as one of the very early views of the concept of holistic view of cosmos. It had a great influence in the formulating the basic postulate of General Relativity. The principle of equivalence of gravitational and inertial mass owes a lot to this principle of Ernest Mach. The generalization of Mach's Principle forms the one of the strong basis of Einstein's theory of General Relativity \cite{Mach}. The moment of inertia of a body is related to the total mass of all the rest of the matter in the Cosmos. But in the case of gravity and inertia it is taken for granted that the influence between any two spatially separated systems cannot be instantaneous. The speed of propagation of gravitational influence cannot exceed the speed of light in vacuum, \textit{c}, satisfying Einstein's strict locality condition. Mach's principle is possibly one of the very early attempts in cosmology to smuggle in the idea of an integrated universe to physics. A suggestion of holism can be read in between the lines. \footnote {A holistic universe is one that is not just determined by the attributes of its individual parts alone but it is governed by the property of the whole, one that has to be considered in totality.}
\section{The New Quantum World View}  
A paradigm shift in the world view has been enforced by the advent of QM. Uncertainty principle is indispensable to QM and is in a way the central pillar supporting QM, if disproved experimentally or contradicted theoretically, the entire existing edifice of QM would just fall apart. The complementary variables of the uncertainty that do not commute can be shown to be Fourier transforms pairs. Planck length and time are defined in terms of the natural units as $\lambda_{p} = \sqrt{Gh/c^3} $ and $t_{p}=  \sqrt{Gh/c^5} $. Here $\lambda_{p}$ and $t_{p}$ are in fact \textit{ipso facto} the distance and time below which the principles of causality and locality may be breached and hence can be treated as units of length and time in classical sense\cite {NaraFourier}. There could be variations to the same \cite{HooftGod}. We interpret the term 'cause' to be transfer of energy from one system to another so that the \textit{law of conservation of energy} becomes almost synonymous with the textit{principle of causality} in the micro quantum realm. Hence any infringement of the principle of conservation of energy at the micro level can be viewed as violation of causality principle. Given that principle of causality is interlinked with the arrow of time\footnote{That cause always precedes the effect is related with the forward arrow of time that we perceive.}, the non-conservation and flouting of the arrow of time happen together. This can be related to Noether's theorem as an inverse case of it.  Heisenberg's energy-time uncertainty principle can also be scrutinized in this context. Both uncertainty in energy and time are sanctioned by this principle. The possibility of Faster Than Light (FTL) propagation amounting to transgression of causality and forward arrow of time is brought about by the situation of non-conservation of energy \footnote {Quantum tunnel effect can be viewed as contravention of conservation of energy for a brief period and consequently the breach of principle of causality on a micro scale}. The upshot of quantum phenomena like tunnel effect effect is a kind of FTL transport as is seen in the recent literature \cite {ZeroTime}. 

A similar logic can be used to relate the momentum conservation law with that of locality. Only a localized particle can transfer momentum. Uncertainty of momentum can lead to the transgression of the sacrosanct classical principle of locality. A similar way the reality of a dynamical variable (say, angular momentum) is at stake due to the Angular momentum angle uncertainty. In a sense uncertainty principle is the main pillar that supports most of the weirdness of QM. Heisenberg's uncertainty principle followed by Max Born's statistical interpretation of the wave function could be considered as the precursor to the so called breakdown of the classical determinism. In short, causality principle of classical physics becomes a casualty in such situations\cite {AchuTime}. Consider the following relations 
of Heisenberg:
\begin{equation}
         \Delta p. \Delta x \geq \hbar / 2 
     \end{equation} 
\begin{equation}   
         \Delta E. \Delta t \geq \hbar / 2  
\end{equation} and 
    \begin{equation}                               
         \Delta J. \Delta \theta \geq \hbar / 2 
\end{equation}
wherein the classical feature of determinism turns out to be at stake. The probability of finding a particle in the interval between \textit{x1 } and \textit{x2} can be expressed using Max Born's exposition of the quantum mechanical wave function as probability amplitude as
\begin{equation}
	p_{x1,x2} = \int^{x1}_{x2}\psi^{*}\psi  dx
\end{equation}
Again, the probabilistic character of physical attributes is encapsulated in the following equation comprising 
\begin{equation}
\left\langle o \right\rangle = \int^{x1}_{x2}\psi^{*}\widehat{O} \psi   dx
\end{equation}
Equations (1) to (5) in a sense characterize the quantum reality of physical attributes. It is  construed that these physical attributes do not possess any definite values or reality unless actually measured or observed in some way. This purported epistemic probabilistic feature of wave function and the measurement problem have given impetus to the conjecture by some noted scientists that consciousness of the observer being responsible for the collapse of wave function or state vector reduction. \textit{Undoubtedly free will finds no place in an utterly deterministic world}. Though QM is very well supported by the numerous experimental results and successful predictions it is not really free from paradoxes a really serious matter calling for further careful study. May be, one may have to look further deeper to salvage the wrecked the deterministic philosophy. The orthodox Copenhagen interpretation stands in confrontation with the classical illustration of reality itself. That reality of dynamical variable or at least its magnitude is created by measurement or observation. That is tantamount to having no reality unless measured. This brings in a plethora of new problems of free will, consciousness and the like \footnote {Substitution of free will completely by 'random generators' is not correct. Random generator can only mimic the unpredictability of free will. Additionally, free will, is capable of creating well defined, organised and meaningful patterns apart from random ones. Both patterns are generated at will which random generators completely fail to.}. 

\section {EPR-Bell Ramifications}   

Let us look at the matter a little closer. Now consider the case of an EPR \cite{EPR} type of experiment. Violations of Bell's inequality \cite{Bell} by EPR type systems have been conclusively demonstrated by Aspects experiment \cite{Aspect} and ample number of experimental results that followed 
These results and standard quantum mechanical elucidation taken together would entail a non-local correlation between two (or more) particles/ components of a system separated by not only time-like but space-like regions as well. The fact is that this is an infringement on the spirit of Special Relativity theory 
even though it is proved that EPR type arrangements cannot be used for transmission of information faster than light. Hence we may very well assume, at least for the present, that such infringements to be evidence of the QNL as has been stated and stressed in the literature profusely. Perhaps in a way Bohr had anticipated the inevitability of such 'weirdness' of QM as evidenced from his reply \cite {Bohr} to EPR and his statement as early as 1927 and from his famous statement\footnote{"Anyone who is not shocked by quantum theory has not understood it"-Neils Bohr.}. \textit{If the hidden variables are ruled out} then the options are only the abandonment of at least any one of the three classically cherished principles-Locality, Reality or Causality. 

\section {Overtones of NL}   

We discuss certain overtones of NL that have far reaching connotations. QNL implies that every region of the space-time is in 'instant' and constant contact with each and every other region of the universe. Of course, the word instantaneous may loose the meaning for space-like separated regions due to relativistic injunction. That implies quantum correlations are not only across space but in time as well. Each element of a system would act according to what states the rest of all other elements of cosmos are in, not just at the present but of the past and future as well. This fact follows from the relativity of simultaneity in time-like regions. It is more serious when the regions are separated by space-like regions. The order of events in one inertial frame could be different in another suitable inertial frame for such space-like separated cases\cite {Emperor}.
In other words QNL indicates that each particle has not only the entire 3-D map but a 4-D map as well (comprising of the 3+1 dimensions of the Minkowski space-time) encompassing both space-like and time-like regions, past and future cones not excluding. The symmetry of past and future gain more significance by the celebrated CPT theorem \cite {AchuFeyn}. This 'information' ought to be updated for every instant of past as well as future, as simultaneity itself is not invariant but frame-dependent according to  special relativity. It signifies that every ostensible individual system has the updated version of the information of the cosmos in its entirety. A particle, for instance ought to be 'cognizant' about the positions of detectors placed at any point in space and time and is apparently 'aware' of whether a certain route is closed or is open. During and shortly after the big bang all the particles in the universe being in a close proximity would have interacted at least once and must hence form components of the EPR type systems. The constantly updated road map of the entire universe should be at the disposal of all particles at every instant of time. Not only spatial correlations but also temporal correlations become imperative here. 

The inevitable logical outcome of the above arguments is that the universe cannot be but \textit{holistic} \cite{IITbell} each element or part is not isolated but depends on the rest of the whole cosmos. An individual particle becomes sort of 'omniscient' in some sense-may be a great deal more! An absolutely isolated system does not exist. The entire cosmos becomes an integrated entity. Facts that we assume to be correct, like our physical laws and constants remain the same across the universe, itself is a suggestion of holistic nature. This may be termed as weak holistic principle. True that such assumptions do not ensure the incessant and instantaneous communication of local changes. QNL would form and provide stronger grounds for the holistic theory. 

\section {Consciousness connection}          

Additionally if the universe were holistic, then it would lead to the consequent crucial proposal -a definite connection between consciousness and the set of entire manifest entities that we justifiably call physical universe. The term consciousness may not overtly find a place in rigorous physics so far. Yet, some of the paradoxical situations arising from QM insist on its inclusion \cite{Tirupati,Agra}. Here, we define the consciousness to be the feeling of 'SELF' in the (higher) organisms, free will being one of its distinguishing and defining features. The solo nature of the self or consciousness and the singleness of the holistic cosmos would robustly require both to be very intimately interrelated. Moreover our daily experience unmistakably demonstrates that our conscious effort (free will?) can produce tangible physical results. One can definitely interact with the physical universe; make changes in the atoms and molecules of our own physical body and hence outside of it as well, by a mere exercise of the will. One may perhaps take a look at Schr\"{o}dinger's argument\cite{Wilber} on this issue though it may not be necessary for the purpose of establishing the point. 
The above arguments clearly lead to our inevitable proposal here-the CCC model. The CCC model is necessitated by the fact that consciousness do interact with a holistic cosmos producing tangible consequences. Both consciousness and cosmos being solitary one may even further conclude them to be inseparable and part of a single entity. Logically follows that they are the same entity in two distinct manifestations. In other words they are two different singular expressions, an internal and an external-the self and the cosmos respectively. Else, consciousness would be a separate entity inside a holistic universe - an extra physical agent capable of affecting the same. The concept of non-duality follows from the above reasoning with the observer occupying an essentially exclusive and fundamentally prominent central position. The physical laws would then turn out to be complementing the psychic laws conveniently. 

Proof of the QNL is only a sufficient but not necessary condition for proving holistic nature of cosmos . Again proof of holistic nature of universe is only enough but not indispensable for proving the idea of CCC. While the above reasoning is, of course, subject to the confirmation of NL, it is not necessarily confined to QNL; the latter becomes an adequate but not an essential condition for the above deduction (viz. that the universe is holistic). Even if QM is proved to be not non-local in character but NL of cosmos is possibly be established the above premises would still turn out to be be true. That is, the idea of CCC would still be the convincing conclusion. Hence the question of NL, whether or not that of QM is of utmost significance. Possibly, it is one of the most urgent of all the epistemological problems of the present situation as important as the question of quantum reality.  Perhaps, that is one of the reasons why Einstein himself considered the question of addressing quantum enigma as the most urgent one that everybody should attempt to immediately. 
\section {Concluding Remarks} 

What was considered the most objective branches of all sciences (physics) is now becoming more and more and more subjective or at any rate omnijective. Words like 'observer' crept in with relativity theories. Terms like 'information','knowledge' though were already important to physics their occurrences are on the increase. In addition, terms like free will and consciousness find all their way to hard core physics. Rightfully it will be more so in future as long as one is concerned with cognition which has much to do with the relation to self. This is regardless of NL in physics is true or not.

We propose that any form of NL not necessarily the quantum version of it, should imply a holistic character of universe. The holistic nature in turn would strongly suggest a major role for consciousness leading to a consciousness Centered / Controlled Cosmos (CCC). In brief we argue that if NL is irrefutably established, it must propel the idea of holistic Universe which in turn would demonstrate the involvement of nothing other than consciousness in scheme of ultimate reality as a whole. It is our firm considered view that consciousness is in fact everything. We intend to present set of further details as to why and how do we arrive at this semi-final conclusion soon. Meanwhile we are attempting to connect up some of the recent concepts and thoughts regarding consciousness and its correlation to NL, as depicted in this paper. 
\begin{acknowledgements}
The authors are highly indebted to our most beloved and Revered \textbf{\textit{Mata Amritanandamayi}} (Chancellor, Amrita Vishwa Vidyapeetham, Deemed University) for giving 
an oportunity to do this work and for the unfailing inspirations. They thank Prof.Dr.Vankatrangan providing support and also thank Prof.Dr.C.S.Shastry and Prof.Dr.N.N.Pillai for useful discussions. 
\end{acknowledgements}


\begin{thebibliography}{}
%
%
\bibitem{Hooft} 
Gerard 't Hooft, 'Quantum Gravity as a Dissipative Deterministic System', arXiv:gr-gc/9903084v3(1999)

\bibitem{HooftGod} 
Gerard 't Hooft, 'How does God Play? (Pre-)Determinism at the Planck Scale', arXiv:hep-th/0104219v1(2001)
 
\bibitem {Unni} 
Unnikrishnan, C. S.: Found. Phys. Lett., 15, 1-25 (2002)

\bibitem{Cramer} 
Cramer, J.G.: Generalized Absorber Theory and the Einstein-Podolsky-Rosen Paradox, Phys. Rev. D. 22 (2), 362-376 (1980)
\bibitem {Pauli} 
 Pauli,W.: General Principles of Quantum Mechanics, (Translated from German by P.Achuthan and K.Venkatesan), Allied publishers, New Delhi, pg.1 (1980) (The original German edition of this classic work was published under the title, Handbuch Der Physik, Band V Teil 1; Prinzipien der Quantentheorie I, Springer-Verlag, Berlin Gottingen Heidelberg (1958)
\bibitem{Stapp} 
Henry. P. Stapp.: Mind, Matter and Quantum Mechanics, Springer, Berlin, Heidelberg, New York,  Preface to first Edition-X (2004)  
 \bibitem{Bohm} 
 Bohm, D.: Wholeness and the Implicate Order, Routledge, London, New York (2002)

\bibitem{HooftDeterm} 
Gerard 't Hooft, 'Determinism Beneath Quantum Mechanics', arXiv:quant-ph/070109v1 (2007)
 
\bibitem{Mach} 
 E. Mach, The Science of Mechanics (Open Court, La Salle, 1960), p. 267, Transl.T.J.McCormack.
   
 \bibitem {NaraFourier} Narayanankutty Karuppath and Achuthan,P.: On Discrete form of FFT, Proc. International conference on Discrete Mathematics and its Applications, Narosa Publ.(2006)

\bibitem{ZeroTime} 
  Gunter Nimtz and Astrid Haibel, Zero Time Space, Wiley-VCH Verlag, Weinheim (2008)

\bibitem{AchuTime} Achuthan,P., Narayanankutty, K., Shastry. C.S.: 'A Brief study of Time', Abstracts of the International Conference on Relativity-(ICR-2005), Amravati University, Maharashtra, 11-14 Jan 2005
\bibitem{EPR} 
 Einstein, A., Podolsky, B., Rosan. N.: Can Quantum-Mechanical Description of Physical Reality  be   Considered Complete?,  Phys. Rev. 47, 777-780 (1935)
 \bibitem{Bell} 
 Bell, J.S.: On the Einstein Podolsky Rosen Paradox, Physics, 1, 195-200 (1964)
 \bibitem{Aspect} 
  Aspect, A., Grangier, P. and Roger, G.: Phys. Rev. Lett. 49, 91-94 (1982) 


\bibitem{Bohr} 
Bohr, N.: Can Quantum-Mechanical Description of Physical Reality be Considered Complete?, Phys. Rev.,48,696 (1935).

\bibitem {Emperor} 
Roger Penrose.: Emperor's New Mind, Oxford Univ.Press, N.Delhi (2005), 370-371 


\bibitem {AchuFeyn} Achuthan,P., Narayanankutty Karuppath, Aspects of Feynman graphs, Electronic notes in Discrete Mathematics, 33, 43-50(2009)

\bibitem{IITbell} 
Narayanankutty Karuppath., Achuthan, P.: On Bell's theorem and inequality, Abstracts of the National Symposium on Mathematical Method and Applications (NSMMA-2005), IIT, Madras, December 22 (2005)
\bibitem{Tirupati} 
Achuthan, P., Narayanankutty Karuppath.: Aspects of Consciousness and Applications, Paper Presented at the 3rd All India Conference on Science and Spiritual Quest (AISSQ-2007), Tirupati, 22-23 Dec, 2007
\bibitem{Agra} 
Achuthan,P., Narayanankutty Karuppath: On Mathematical Modeling of Quantum Mechanical Systems, Paper at the International Conference on Modeling of Engineering and Technological Problems (ICMETP) and 9th Biennial Conference of Indian Society of Industrial and Applied Mathematics, on , held at BMAS Engg. College, Agra, India. 14th -16th Jan, (2009)  

\bibitem {Wilber} 
Schr\"{o}dinger, E.: Quantum Questions, pp. 92-93 Eds. K.Wilber, Shambala, Boston, (2001)


\end{thebibliography}


\label{references}                    

\end{document}